# Significance of Ultra High Energy Cosmic Rays (UHECR)

Matter is generally found to be in thermal equilibrium everywhere in our Universe. However, there is matter which deviates extremely from this thermal equilibrium. The corresponding energy limit of the particles associated with this form of matter extends from 1 TeV to beyond $10^{20}$ eV. UHECR have energies above $10^{20}$ eV. The reason and the process by which these particles gain such remarkably high energies are mysterious. These particles may lead to the discovery of unknown physics or of unknown exotic particles found in the early universe.[1] In fact anti matter, muons, and neutrinos were all discovered while studying cosmic rays. Therefore, we can expect UHECR`s to provide big breakthroughs in our study of dark matter, dark energy and the structure and composition of our Universe. It may also give us important insights into the nature of gravity as well as the existence of extra dimensions [2]. The OWL mission at NASA studies these additional important questions related to UHECR. For more details on the importance of UHECR`S see [1] and [2].

It should be noted that the energy required to accelerate a particle to $10^{20}$ eV is extremely large. Therefore, the source of these particles will have extremely large energy. One can only imagine how gigantic the source could be, if one considers that the energies of the particles accelerated to $10^{15}$ eV are associated with a gigantic supernova. Clearly, above $10^{20}$ eV there is an extremely gigantic but unknown source which is sending us these extragalactic particles. According to David Schramm, "[UHECR] is a remarkable field in which the most conservation explanation involves super massive black holes" [2]

It is important to distinguish UHECR from Cosmic Microwave Background Radiation (CMBR). Cosmic Microwave Background Radiation is a kind of electromagnetic radiation that is believed to fill out the entire Universe [3]. It is a well known fact that the space between galaxies is black as seen from an optical telescope. However, using radio telescopes when we study this empty space, we find that in the microwave region of the spectrum there is a faint glow. Surprisingly, this glow does not come from any star, galaxy or any other object. It is notable that Cosmic Microwave Background Radiation is the most conclusive evidence for the Big Bang [4]

## A Surprising but important Twist

While CMBR is a leftover radiation from the Big Bang, UHECR are subatomic particles moving at 99.99999999999% of the speed of light. Basically UHECR is composed of matter mainly nuclei. Now, the interesting feature of cosmic rays is that they interact with CMBR. More precisely, it was experimentally verified that protons with energies above $4*10^{19}$ eV interact with CMBR [5] [6]. One of the important predictions of this interaction was the occurrence of a steepening of the energy spectrum between 4 and $10*10^{19}$ eV. This is the famous GZK cut-off. According to this cut-off we should absolutely not see any particles above the range of $5*10^{19}$ eV. The NASA`s Owl Mission which studies these cosmic rays described it as follows:"What is so perplexing is that, from what we understand about physics and the Universe, we shouldn't be seeing many cosmic rays above about $5 \times 10^{19}$ eV, yet we do"[2]. As mentioned earlier these rays with energies above the range of $10^{20}$ eV are known as UHECR`s. They are so rare that they only hit the earth`s atmosphere at the rate of 1km per square per century. Measuring the energy spectrum, the mass composition, and the arrival direction distribution of UHECR`s is of particular significance. One of the most important techniques used in the detection of UHECR`s is Cherenkov radiation.

A brief history of origin of Cherenkov radiation

In the study of early radioactivity, a faint visible light was observed from strong sources of beta and gamma radiation. This bluish-white light was particularly evident in a liquid source. Earlier, it was believed that this radiation was due to ionization caused by beta or gamma particles. However, it was soon observed that the spectrum of radiation was continuous which along with other evidences raised doubts about ionization as a cause of radiation [7]. The Russian Physicist Cherenkov in a series of experiments proved that the light produced by fast charged particles moving through transparent medium has nothing to do with ionization. Cherenkov discovered that light has unique directional and polarization properties.

An elementary theory of Cherenkov Effect was developed according to which electrons are displaced when the particle moving at a uniform velocity in a dielectric medium passes through the associated electromagnetic field. Basically the field polarizes the medium, and the displaced electrons follow the waveform of the pulse. These displaced electrons produce Cherenkov radiation. When the particle is moving slow, no radiation is produced because of destructive interference. When the particle`s velocity in medium is faster than the velocity of light, radiation is produced because the wavelets are in phase. A relationship between velocity of the particle, refractive index of the medium, and the angle at which light is emitted was discovered which is given by:

$$\cos\theta = 1/\beta n$$

For a detailed mathematical treatment of the whole theory see [7] and the references in [7].

Applications of Cherenkov radiation to detect particles

With the help of Cherenkov Effect we can detect and count the single charged particles at very fast rate. These charged particles are at energies in excess of Cherenkov threshold. By determining the velocity of the particle we can also determine the energy of the particle provided the mass is known. When the particles have same energy we can distinguish particles of different masses. The directional feature of Cherenkov radiation enables us to determine the sense in which particle is travelling. In general, Cherenkov detectors are of two types focusing and non-focusing. In the non-focusing type detector light is allowed to come from all directions of the path of the particle. A matt white scattering material of high reflectivity such as magnesium oxide powder is used. The photomultiplier is inserted somewhere inside and it collects a fraction of the total light produced.

In the focusing type Cherenkov detector light coming from one angle falls on a small area. This area is determined by the detector if light has to fall at Cherenkov angle. The difference between light falling at Cherenkov angle and the light at all other angles is sharply contrasted. An example of the focusing detector is shown in fig. 3. In the figure, A represents the brass tube for positioning radiator. B is a Lucite radiator lends. Basically, radiator and lens are combined in a single cell. C is a cylindrical mirror. D is a photomultiplier tube of type 1p28. E represents plane mirrors for splitting image.

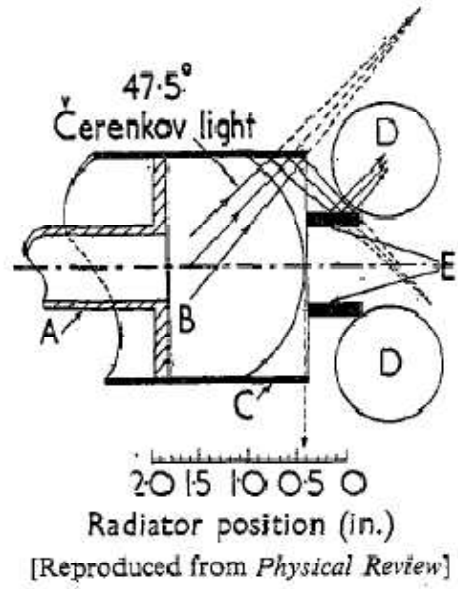

Fig. 3. One of the counters designed and built by Marshall[14] having focusing properties

Figure 1, is reproduced from the article written in 1955 by one of the pioneers in the work on Cherenkov radiation J.VJelley. For a detailed description see [7].

A ring of large radius is the focused image. The role of the cylindrical mirror is to gather the light into a smaller area. Hence, it is very close to radiator in order to focus the light near optic axis. The image is split by two small plane mirrors. Light gets divided between two photomultipliers which are connected in a fast coincidence circuit. Now the rate of pulses from the tubes can be distinguished with the rate of pulses from the particles in the beam with the help of coincident arrangement. The arrangement is such that phototubes are to the side of the main beam of particles. Rays of different Cherenkov angles can be focused on the cathodes of the phototubes by adjusting the position of the radiator in the cylindrical mirror.

Figure 4 shows a standard result obtained by the apparatus in figure 3 for the case of pi-mesons. The energy of a beam of pi-mesons is measured. The taller curve represents 141 Mev pi-mesons while the smaller curve represents 121Mev pi-mesons.

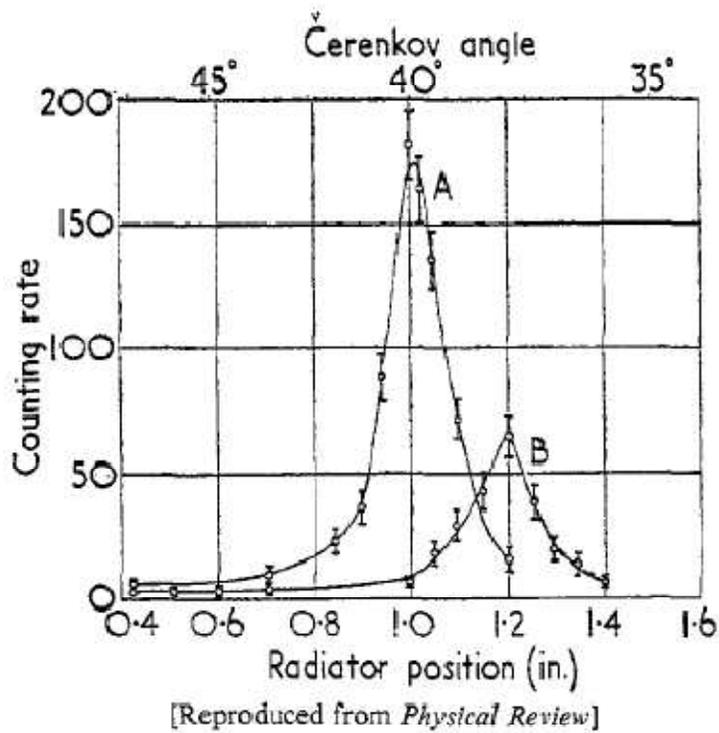

Figure 2, reproduced from [7]

Pierre Auger Observatory

    The cosmic ray particles produce Cherenkov radiation which constitutes about 0.0004% of the light of the night sky [8]. Now, it is very hard to distinguish the light corresponding to these particles from the average light intensity of the night sky. However, by studying air showers produced by Ultra high energy cosmic rays and by using the pulse technique we can measure individual light pulses. Hence, we can also measure individual air showers. These air showers are basically composed of billions of secondary particles that are produced when ultra high energy cosmic rays strike the atmosphere [9]. Because of the rarity of the particles striking the

Earth above the range of $10^{20}$ eV, a truly giant detector is needed. The Pierre Auger Observatory was designed to suit this purpose.

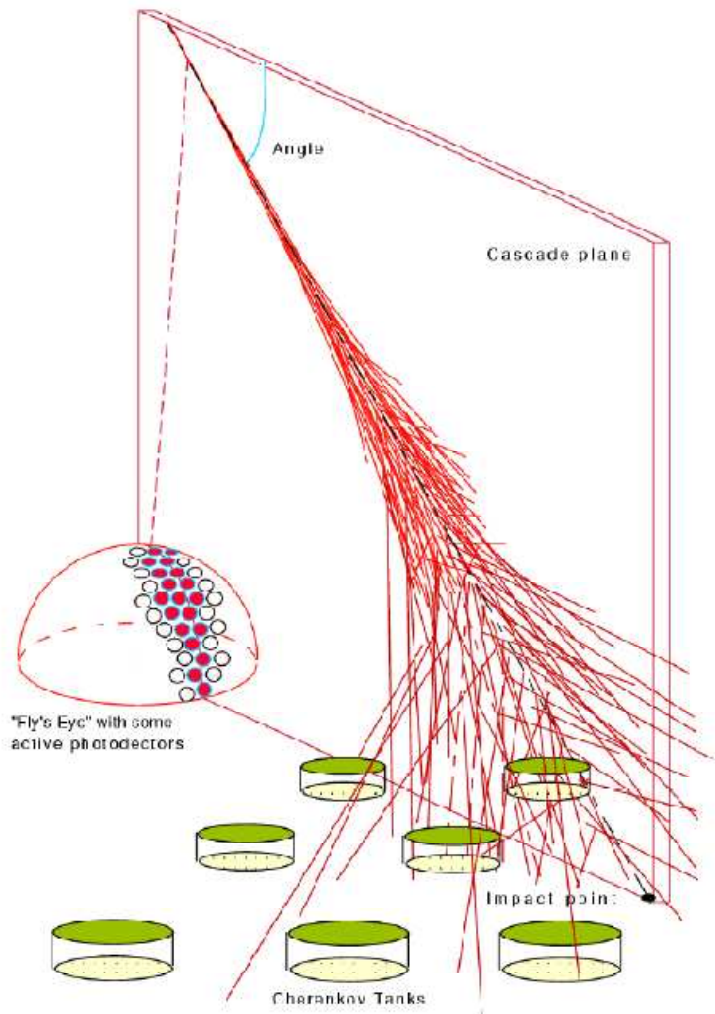

Figure 3: A schematic illustration of the hybrid concept of the Auger Observatory [10]

The observatory uses a hybrid system of four fluorescence detectors and surface detectors (see fig.3). The fluorescence detectors operate only at night and usually work well best on moon less nights. The tangential composition of the shower is measured at the surface level by the ground array which also helps in distinguishing the electromagnetic and muon components [11].

At the atmospheric level, the fluorescence detectors measure the longitudinal compositions of the showers. [11]

The fluorescence detectors contain six identical fluorescence telescopes. A 1.1 m radius diaphragm allows fluorescence light to enter the telescope. Then a spherical mirror of area 3.5*3.5 m$^2$ collects the light which is then focused on a photomultiplier tube camera. The camera contains 440 hexagonal PMTs of diameter 45mm. Each PMT covers 1.5$^0$ diameter of the fraction of the total sky. The optical spot size on the focal surface has a diameter of 15mm for incident light from all directions. Small light reflectors are used in order to reduce signal losses when spot light passes PMT boundaries. The field of view of a single telescope is 30$^0$ by 28.6$^0$. For more details and for a firsthand account on the working of Fluorescence detectors see [11]. See Figure 4 for a schematic view of the Fluorescence detector telescope.

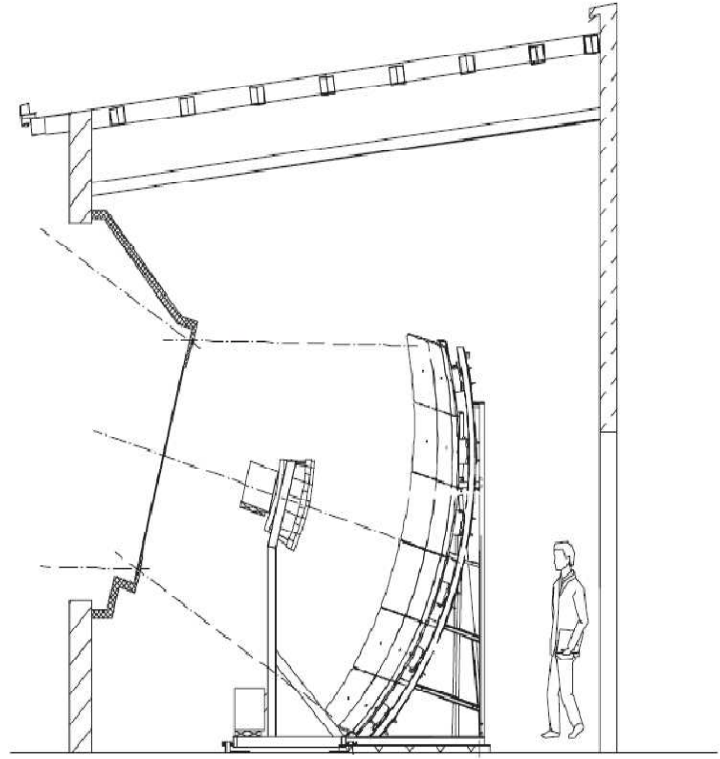

Fig. 6. Schematic view of a fluorescence detector telescope. From left to right can be seen the aperture system, the PMT camera and the spherical mirror.

Figure 4, taken directly from Nuclear Instruments and Methods in Physical Research [12].

The second component of the hybrid system used to detect UHECR`s are the surface detectors. The surface detectors are spread out over an area of 3000 km$^2$ in the form of a triangular grid. Every site in the Pierre Auger Observatory contains 1600 water Cherenkov detectors with a spacing of about 1.5 km. These detectors contain about 12 tons of water. There are numerous advantages of using water Cherenkov detectors as surface detectors. They are relatively cheaper than scintillators Secondly; they can detect large number of showers at large zenith angles, and they respond well to the particles.

Water is filled in the tank at the height of 1.2 m. Cherenkov light emitted by particles which pass through the tank is collected with the help of three 9" Photonis XP 1805 photomultiplier tubes. Signal is received both from the anode and the last dynode. The dynode is amplified so that it can achieve a range from a few to $10^5$ photoelectrons. A 5 pole anti-aliasing filter of cut-off frequency 20 MHz connects all the channels. The filter is digitized at 40 MHz by 10 bit FADC.A local CPU operates the digital trigger. A GPS takes care of time while all communications to Central Data Acquisition System takes place by using a specially designed wireless communication system. There are two solar panels which provide 10W of power to the electronics by charging two 12 V batteries. Every detector can operate independently of all other detectors. (See figure 5 in the next page). For more detailed description of the functioning of the surface detector see [12] and [13].

Let us consider one simple example of how data is collected using these hybrid detectors. In figure 6, data is collected for an event that arrives at $34^0$ from normal. There were 14 detectors which were above trigger level. The shower was produced by high energy cosmic rays with the energy estimated at ~$10^{19}$ eV.

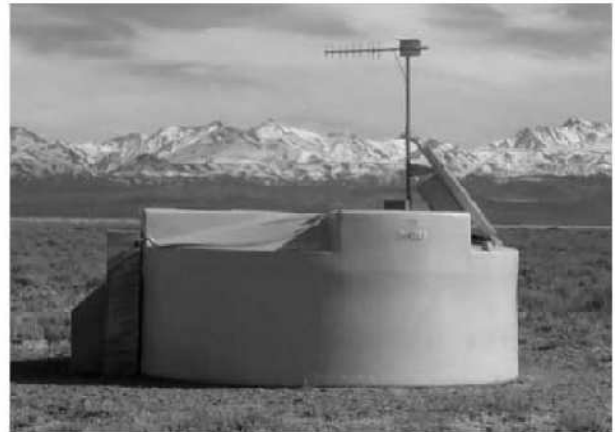
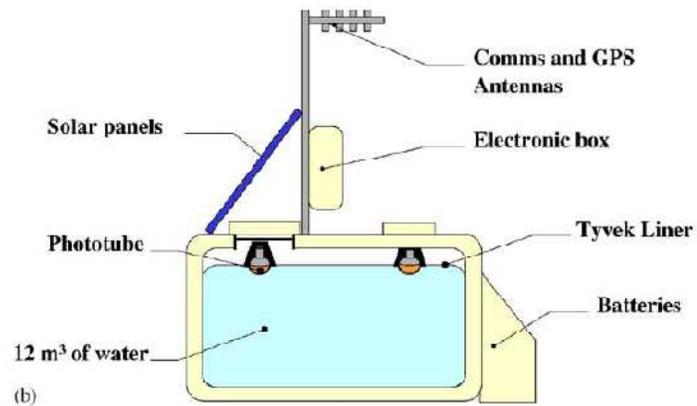
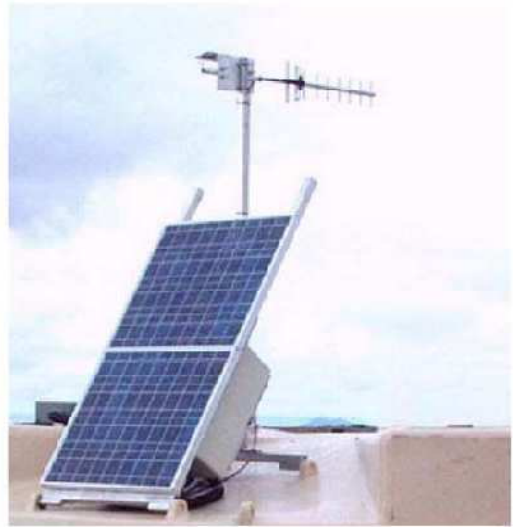

Fig. 2. (a) A photograph of an EA water tank; (b) schematic view of an EA tank; (c) the Yagi antenna and the solar power array.

Fig. 5, taken directly from Auger Collaboration published in Nuclear Instruments and Methods in Physics research [12].

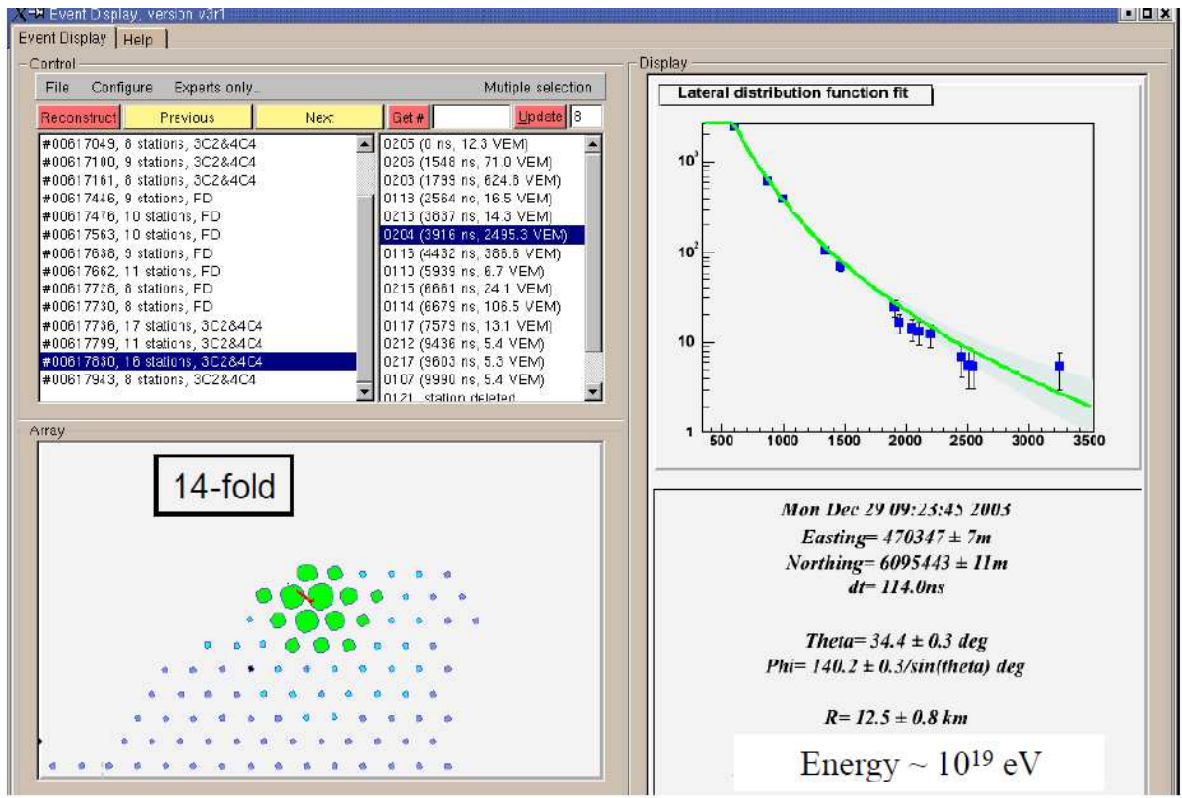

Figure 5. A 14-fold event at 34° from the zenith. The figure shows the on-line event display. The sequence of events is shown in the top left corner with the density pattern of the triggered detectors below. The lateral distribution and some parameters of the event are in the right hand panels.

Figure 6 taken from [10]

Figure 7 in the next page shows the data for an event that triggered 32 detectors separated by over 15 km. The data arrived at an angle of $72^0$ from normal. From the radius of curvature and FADC traces it was estimated that the event was created by Baryonic rays. Energy was found to be above ~ $10^{19}$ eV.

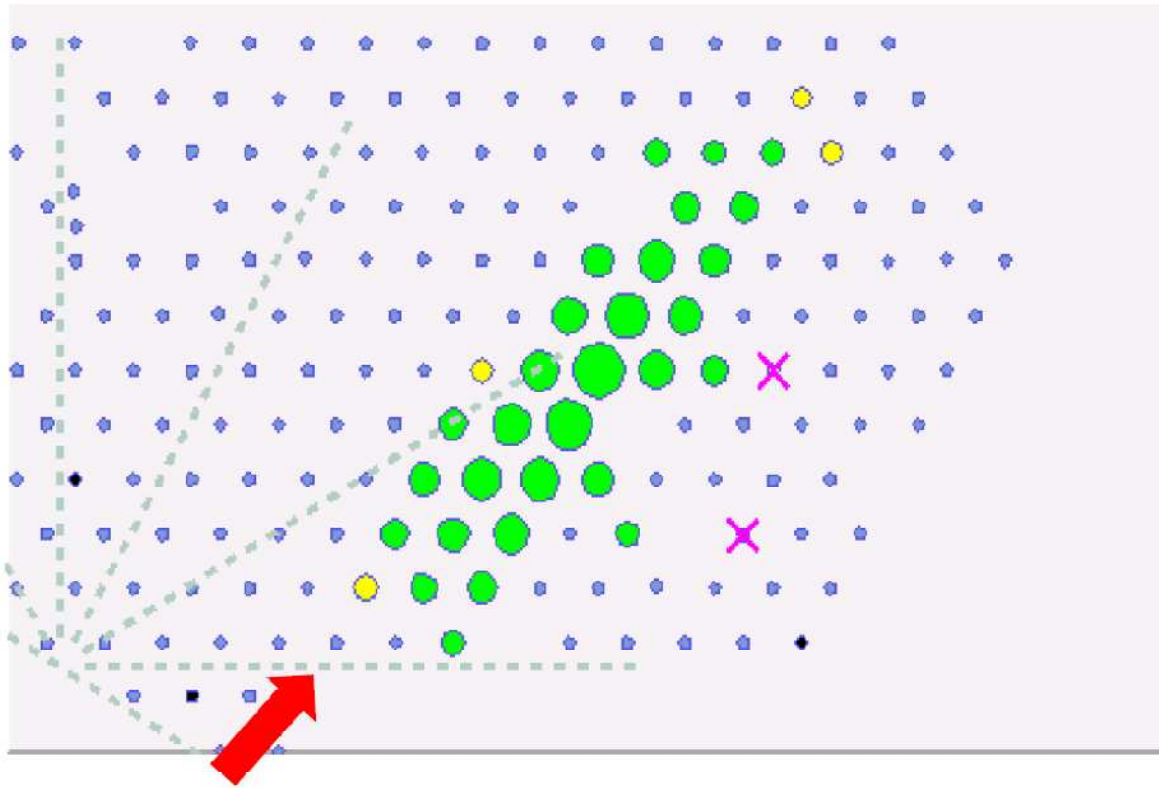

Figure 6: An event with 33 tanks triggered. The zenith angle was 72° and the radius of curvature was 37 km. The energy of the primary was > 10$^{19}$ eV.

Figure 7 taken from [10]

Conclusions

As established in the introduction, Ultra high energy cosmic rays are very significant in our understanding of the mysteries of the universe. These rays upon entering the earth`s atmosphere produce billions of sub atomic particles and form air showers. By studying the properties of these showers like their spectrum, mass composition etc. we try to know more about the source of these rays. Since these particles come at very high speeds they emit Cherenkov radiation. Because of the rarity of these events we have to use extremely large detectors to measure them. The Pierre Auger Observatory uses hybrid detectors. The surface detectors used are water Cherenkov detectors about 30 times the size of the city of Paris. Investigations are on, and we should expect to get more insights into the source of these rays in future.